\documentclass[]{uiuc_paper}
\pdfoutput=1

\usepackage{cleveref}
\crefname{figure}{Figure}{Figures}
\Crefname{figure}{Figure}{Figures}
\crefname{table}{Table}{Tables}
\Crefname{table}{Table}{Tables}
\crefname{section}{Section}{Sections}
\Crefname{section}{Section}{Sections}
\crefname{subsection}{Section}{Sections}
\Crefname{subsection}{Section}{Sections}
\crefname{equation}{Eq.}{Eqs.}
\Crefname{equation}{Equation}{Equations}
\usepackage{booktabs}
\usepackage{tabularx}
\usepackage{amsmath}
\usepackage{amssymb}
\usepackage{fix-cm}
\usepackage{lmodern}
\usepackage[T1]{fontenc}

\usepackage{url}
\usepackage{amsfonts}
\usepackage[utf8]{inputenc}
\usepackage{multirow}
\usepackage{makecell}
\usepackage{xspace}
\usepackage{adjustbox}

\newcommand{\eg}{\textit{e.g.}\xspace}

\newcommand{\dq}[1]{``#1''}

\newcommand{\rw}[1]{\textsf{\small #1}}
\newcommand{\niaolei}{ni\v{a}ol\`ei}
\newcommand{\zongtong}{z\v{o}ngt\v{o}ng}
\newcommand{\yinyue}{y\=inyu\`e}
\newcommand{\gundong}{g\v{u}nd\`ong}
\newcommand{\huaxue}{hu\'axu\v{e}}
\newcommand{\laohu}{l\v{a}oh\v{u}}
\newcommand{\bian}{bi\=an}

\newtcolorbox{findingbox}[1][]{
  enhanced,
  colback=orange!8,
  colframe=orange!30,
  fonttitle=\bfseries,
  coltitle=black,
  colbacktitle=orange!45,
  title=#1,
  left=4mm, right=4mm, top=1mm, bottom=2mm,
  boxrule=0.8pt,
  arc=3mm,
  toptitle=2mm, bottomtitle=1mm
}

\title{Can We Read the Mind of an Audio LLM?\\ A Verbalizable, Multilingual Middle-Layer Workspace}

\author[1,2]{Jiajun Fan}
\author[1]{Jingyuan Li}
\author[1]{Prashanth Gurunath Shivakumar}
\author[1]{Qi Luo}
\author[1]{Jia-Hong Huang}
\author[1]{M. Maruf}
\author[1]{Roger Ren}
\author[1]{Yile Gu}
\author[1]{Rahul Pandey}
\author[2]{Ge Liu}
\author[1]{Ivan Bulyko}

\affiliation[1]{Amazon AGI Foundations}
\affiliation[2]{University of Illinois Urbana-Champaign}

\abstract{An audio language model is a black box in a specific way: we see what it says, never what it works out on the way there, and chain-of-thought monitoring helps only if the model writes its reasoning down. Reading a base Qwen3-Omni with a logit lens at the audio-token positions, we find that the answer to a spoken question becomes legible---in words---in the model's middle layers, before it emits any token. Five findings follow. (1) The readout carries concepts in neither the question, the options, nor the model's own transcription: on a clip whose verbatim transcription is empty garbling, it reconstructs \emph{Watergate} and \emph{scandal}, passes through the role \emph{president}, and resolves to \emph{Nixon}---a hidden multi-hop chain, read with no chain-of-thought. (2) The content is language-agnostic: one audio-inferred concept surfaces in several scripts at once, and 38\% of top-1 readouts are Chinese on English inputs. (3) It is paralinguistic: given the same clip as audio and as the model's own emotion-free caption, the audio mind forms the sound source, speaker role, or affect that the caption discards, and answers correctly more often. (4) The audio-driven signal is absent at the input, turns on about a tenth of the way into the network, separates most cleanly from the text prior in the middle band (35--80\% of depth), and activation patching shows it is causally used and committed before the last fifth of the layers. (5) Deleting single layers maps the pipeline: reading the sound in is localized to the entry layers and answer delivery to the output layer, while retrieval is distributed across the interior. Throughout, a waveform-swap control---identical text, only the sound changed---isolates the audio-driven signal from a prior over the printed options. This is a qualitative account of what an audio model works out before it speaks: the quantities are controls, not benchmark scores.
}

\date{\today}

\begin{document}

\maketitle

\section{Introduction}

An audio language model has been a black box: we see what it says, never what it
thinks. Chain-of-thought helps only if the model writes its reasoning out---and a
model that answers a spoken question directly writes nothing down. In this work we
read a base audio LLM (Qwen3-Omni) during the forward pass, as it listens to a spoken
question and before it answers, and watch intermediate concepts---the source of a
sound, a speaker's role, an inferred event---take shape in its middle layers and
become legible as words, though they appear in neither the question nor the answer.

What makes this possible is that a model's intermediate computation is, to a
meaningful degree, legible. By projecting intermediate activations into vocabulary
space, the logit lens \citep{nostalgebraist2020logitlens} and its calibrated successors
\citep{belrose2023tunedlens} surface partially formed predictions well before those
tokens are produced. Building on this, \citet{gurnee2026workspace} argue that these
readable representations are not incidental: language models maintain a small,
privileged set of \emph{verbalizable} representations concentrated in the middle
layers that is used for internal reasoning, broadcast onward, and selectively
recruited---a functional global workspace \citep{baars1988cognitive,dehaene2001workspace}. Their
analysis exposes three functional bands: \emph{sensory} early layers, a
\emph{workspace} band in the middle ($\sim$35--80\% of depth) where verbalizable
concepts are represented, and \emph{motor} late layers that align to the output.
Those results were established for text, code, and image inputs; whether the
structure extends to other modalities was explicitly left open. Audio was not tested.

This matters beyond curiosity. Speech agents consume raw audio, may reason over
paralinguistic content a transcript discards, and can emit speech or tool calls
directly; if their decision-relevant concepts were legible before they act, that
would be a lever for interpretability-based oversight---precisely the regime a
chain-of-thought monitor \citep{baker2025monitoring,korbak2025chain} cannot see, since
stated reasoning need not be produced and can be unfaithful
\citep{turpin2023language,lanham2023measuring}. But the workspace picture was established only for
token inputs. When an audio LLM listens to a spoken question, does it form a
middle-layer, verbalizable structure in which the answer---derived from the
sound---can be read out, or is any readable structure merely a text prior over the
accompanying prompt?

We study base Qwen3-Omni-30B-A3B-Instruct \citep{qwen3omni}, whose audio encoder
injects features at audio-token positions into a 48-layer text \emph{Thinker}. We
apply a logit lens at those positions and read, for each answer option, the best rank
achieved by its first token. Our central control fixes the question and the options
byte-identical and swaps only the waveform (the real clip, a mismatched clip, or
silence), which separates a genuinely sound-driven readout from a prior over the
printed text. We organize the evidence as five findings.

\begin{enumerate}\setlength{\itemsep}{1pt}\setlength{\parskip}{0pt}
\item \textbf{We can read the model's mind, with no chain-of-thought}
  (\cref{sec:f1}): on a clip whose verbatim transcription is empty garbling, the
  readout assembles \rw{water}$+$\rw{gate}, then \rw{scandal}, then the role
  \rw{president}, and resolves to \rw{Nixon}---a hidden multi-hop chain.
\item \textbf{The thinking space is multilingual} (\cref{sec:f2}): the same
  audio-inferred concept surfaces in several languages at once, and $38.5\%$ of top-1
  readouts are Chinese on entirely English inputs.
\item \textbf{When speech carries extra information, the audio mind registers it and
  the text mind does not} (\cref{sec:f3}): given the same clip as audio and as the
  model's own emotion-free caption, only the audio mind forms the true sound source,
  speaker role, or affect.
\item \textbf{The thought lives in the middle and forms in depth order}
  (\cref{sec:f4}): the audio-driven gap over the text prior is null at the input,
  significant from $\sim$12\% depth, most cleanly separable from the prior in the
  workspace band, and causally used before the motor layers.
\item \textbf{The answer is distributed, and hallucinations are born early}
  (\cref{sec:f5}): deleting layers localizes listening to the entry layers and
  delivery to the output layer, while no interior layer is essential; a text-mode
  hallucination usually takes over early and is never revised.
\end{enumerate}

This is a qualitative account, and we are deliberately conservative about what it
shows: a logit lens rather than the more faithful Jacobian lens, one model family, a
small sound-dominated slice, and a readout that is sparse across audio positions. The
quantities we report are controls---they exist to rule out a text prior, not to score
the model on a benchmark.

\section{Related Work}

\paragraph{Reading transformer internals.} The logit lens \citep{nostalgebraist2020logitlens}
and tuned lens \citep{belrose2023tunedlens} decode intermediate states into vocabulary
space; circuit- and feature-level work \citep{elhage2021framework,templeton2024scaling,
lindsey2025biology} studies how such content is computed.
\citet{gurnee2026workspace} characterize a verbalizable middle-layer global
workspace---verbalizable, reused, broadcast, and gated representations---and introduce
the Jacobian lens, a faithful alternative to the logit lens, to map it. Reading such
internal state is of interest for oversight because chain-of-thought monitors
\citep{baker2025monitoring,korbak2025chain} depend on a model's stated reasoning, which
need not be produced and can be unfaithful \citep{turpin2023language,lanham2023measuring}.

\paragraph{Latent language and audio lenses.} Text models represent meaning in a
language-agnostic ``semantic hub'' \citep{wu2025semantichub,wendler2024llamas}. Closest to
us, two lines already apply a vocabulary projection to audio inputs: the Semantic Hub
Hypothesis \citep{wu2025semantichub} shows audio content is lens-readable in the
model's dominant language, and AudioLens \citep{yang2025audiolens} tracks how surface
auditory attributes (\eg gender, emotion) evolve across layers. Our question is
different in two ways. First, we ask not whether \emph{any} audio content is
lens-readable, but whether the particular global-workspace organization of
\citet{gurnee2026workspace}---a verbalizable middle band carrying task-relevant,
\emph{inferred} content---appears for audio. Second, we add a waveform-swap control
that isolates the audio-driven signal from a text prior over the printed options;
without such a control a lens readout cannot be attributed to the sound rather than
to the accompanying prompt. To our knowledge this control, and the resulting
localization of an audio-driven, task-relevant signal to the workspace band, are new.

\paragraph{What speech models are trained to do, and why we read a base one.} The
readout we study sits under a fast-moving post-training stack. Rewards defined over the
reasoning \emph{process} improve the consistency and scalability of audio-LLM reasoning
\citep{fancesar}; procedure-aware reinforcement learning does the same for
tool-augmented models \citep{zheng2026procedureaware}; and executable pipelines push
multi-hop reasoning out of the activations and into code the system runs
\citep{pyrag}. All three make reasoning \emph{more} externally visible. We take the
opposite starting point---a base model, answering directly, with nothing written
down---and ask what is already legible inside it before any of that machinery is added.

\section{Method}
\label{sec:method}

\paragraph{Model and lens.} We read the Thinker of Qwen3-Omni-30B-A3B-Instruct
\citep{qwen3omni}: $L{=}48$ layers, width $d{=}2048$, vocabulary
$|V|{=}152{,}064$, untied output embedding $W_U$. The logit lens
\citep{nostalgebraist2020logitlens} decodes an intermediate residual state $h_p^{(\ell)}$ by
applying the terminal read-out prematurely,
\begin{equation}
\mathrm{lens}\big(h_p^{(\ell)}\big) = \mathrm{softmax}\big(W_U\,\mathrm{norm}(h_p^{(\ell)})\big).
\label{eq:lens}
\end{equation}
For a target token $t$ we take its rank under this distribution; rank 1 means $t$ is
the top token. Ranks are comparable across layers, where raw probabilities and logit
scale drift with depth. We read at audio-token positions and report the best
(minimum) rank over those positions and over the workspace-band layers (Thinker
layers 17--38, $\sim$35--80\% depth, taken \emph{a priori} from the text-model
account of \citet{gurnee2026workspace}, not tuned on our data); the four-way
prediction is the option of minimum rank. The logit lens is a cheap proxy rather than
the Jacobian lens of \citet{gurnee2026workspace}; relative to a faithful lens our
readouts should, if anything, under-detect, and early-layer readouts are noisy by
construction, so our claims concern the middle band.

\paragraph{Audio-swap control.} Audio is injected at audio-pad positions, which
precede the question in the causal stream, so an audio-position readout cannot attend
to the trailing options. We hold the written question and every option token
byte-identical and replace only the waveform with (a) the real clip, (b) a mismatched
clip of the same category, or (c) silence. This isolates an audio-driven readout from
a text prior over the printed options. Because option positions are imbalanced (an
always-majority baseline scores $\sim$46\%, not 25\%), our primary metrics are
balanced accuracy and the paired McNemar test.

\paragraph{Two-mind comparison.} To ask what the sound adds over its words, we feed
the same clip twice---as the real waveform, and as the model's own emotion-free
caption of it supplied as text with no waveform---with the identical question and
options in both runs.

\paragraph{Causal tests.} Decodability need not mean use. We test use with activation
patching: on clips the model answers correctly with audio and incorrectly once the
audio's mel-features are zeroed, we copy the clean run's residual stream into the
corrupted run at one band of layers and at the audio-token positions, and let the
model finish. Separately, we delete one layer at a time (replacing its output with
its input, an identity skip) and measure the resulting accuracy drop; single-clip
answers are noisy under bf16 mixture-of-experts routing, so we average over clips.

\paragraph{Data.} Core controls run on a curated 140-clip MMAU \citep{mmau2024}
set (non-counting answers whose first token is uniquely discriminative, an 83:57
sound:speech split); corpus-level statistics draw on larger MMAU pools as noted (338
music-track clips, 948 brain-map grids---one grid is one clip's
layer${\times}$position readout, 863{,}770 workspace cells---the 1000-clip MMAU-mini,
55 emotion clips), and on a 500-question spoken TriviaQA \citep{joshi2017triviaqa}.

\section{Findings}

\begin{figure}[t]
\centering
\includegraphics[width=0.82\textwidth]{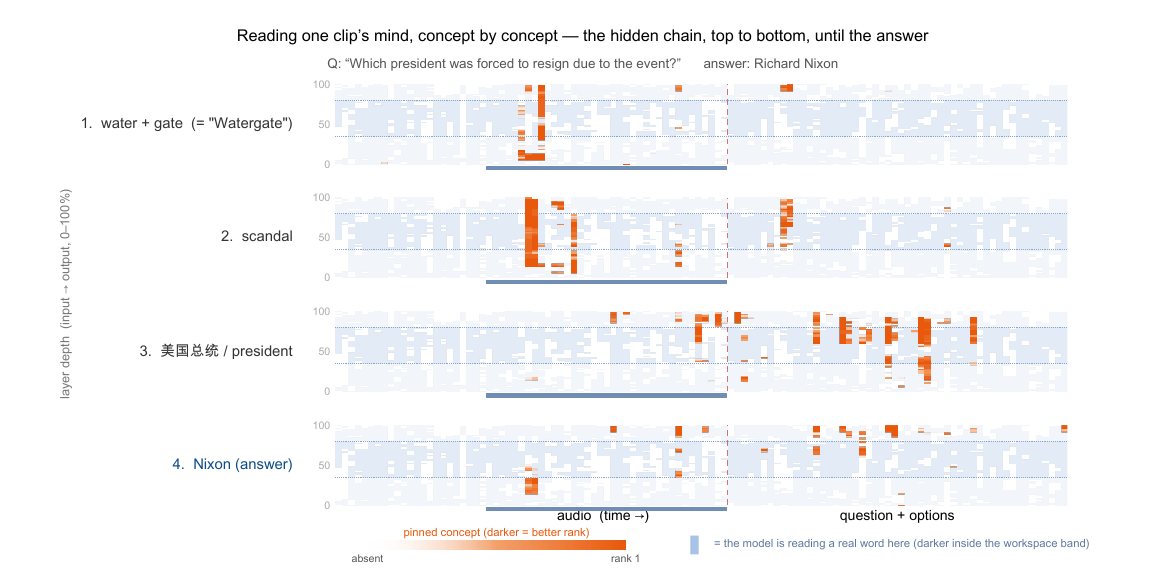}
\caption{\textbf{Reading one clip's mind, concept by concept} (Nixon clip). One row
per concept word; each is that word's logit-lens rank over the full input ($x$: audio
positions, then question${+}$options) against layer depth ($y$). Darker $=$ better
rank; the shaded band is the workspace. The event concepts (\rw{water}/\rw{gate},
\rw{scandal}) are rank 1 from the early layers and the actor concepts
(\rw{president}, \rw{Nixon}) strengthen deeper---an event-then-actor progression
rather than a strict token-by-token order; the answer is reached over the audio, not
only at the printed option.}
\label{fig:nixon}
\end{figure}

\subsection{Finding 1: We Can Read the Mind---No Chain-of-Thought Needed}
\label{sec:f1}

Reading the forward pass at the audio positions, we can watch the model's train of
thought---not just what it answers but how it gets there. The most striking cases are
multi-hop: the workspace surfaces a concept in neither the question nor any option,
inferred from the sound, and only then resolves to the answer.

\paragraph{The Nixon clip (flagship).} Our cleanest case asks which president was
forced to resign (answer: Nixon). What makes it clean is that the model's own two
routes for turning this audio into text both fail to name the chain. Its verbatim
transcription is empty garbling (\dq{The ote --- [End of Audio]}); its free-form
caption recovers only the generic political scene (\dq{\dots president of\dots
political officer\dots politician\dots senator\dots}), never \emph{Watergate},
\emph{scandal}, or \emph{Nixon}. Yet the middle-layer readout assembles exactly that
chain (\cref{fig:nixon}). At adjacent audio positions it reaches rank 1 on \rw{water}
and \rw{gate}---reconstructing \dq{Watergate}, which never appears as a single
token---while \rw{scandal} is rank 1 from the earliest layers. The event so
characterized, the person and the role surface together as the readout deepens:
\rw{Nixon} and \rw{president} both reach rank 1 across the sensory--workspace
transition, with the role persisting in the workspace band as Chinese \zongtong. We
therefore read the chain as event-then-actor rather than a strict token-by-token
ordering. \emph{Nixon} is a printed option, but it is decoded over the audio
positions, not merely read off the choice---this is the one clip where the audio-side
readout reaches the answer name. That two independent audio-to-text routes miss these
concepts makes plain re-transcription unlikely; a failed ASR does not prove the words
were never spoken, so we bound that residual possibility as a limitation.

\begin{figure}[t]
\centering
\includegraphics[width=0.84\textwidth]{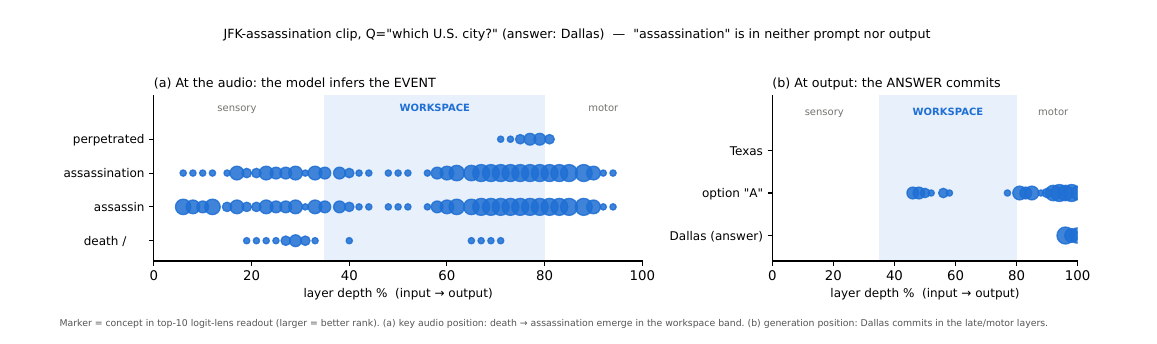}
\caption{\textbf{Reading an audio LLM's workspace before it speaks} (Kennedy clip; Q:
\dq{which U.S.\ city?}; answer: Dallas; larger marker $=$ better rank). \textbf{(a)}
At an audio position the event concept \rw{assassination} stabilizes at rank 1 across
the workspace band alongside \rw{assassin} and \rw{perpetrated}. It appears in
neither the printed question nor the answer, but the model's own ASR partly
transcribes it, so we treat this clip as illustrative rather than pure inference; the
beyond-transcript readout here is the late victim name \rw{Kennedy}
(\cref{sec:f1}). \textbf{(b)} At the generation position the answer \emph{Dallas}
commits only in the motor layers: concept in the middle band, answer at the output.}
\label{fig:chain}
\end{figure}

\paragraph{The Kennedy clip (a partial case, stated honestly).} A clip narrating the
Kennedy assassination (answer: Dallas) is weaker, and we include it to mark the
boundary (\cref{fig:chain}). Here the model's ASR partly succeeds (it transcribes
\dq{the assassination of President John F.}), so \rw{assassination} and
\rw{president} are largely transcribed, not inferred, and we do not count them. What
goes beyond the transcript is the late naming of the victim \rw{Kennedy} (absent from
prompt and answer); the answer \emph{Dallas} is read from the printed option. This
partly-transcription-driven chain, in contrast to Nixon's fully-inferred one, is how
we scope the \dq{read the mind} claim honestly.

\paragraph{The pattern recurs across regimes.} The flagship is one clip, but the
readout structure is not unique to it: \cref{tab:cases} collects seven clips, curated
to illustrate the range, for which the answer concept is read at rank 1 in the
workspace band over the audio positions before any output token. (The
population-level claim is the waveform-swap and band statistics of \cref{sec:f4}, not
these hand-picked cases.) They span three regimes. In \emph{multi-hop inference} an
intermediate concept absent from the prompt is surfaced over the audio and only then
resolved: besides Dallas, an archaeology clip surfaces \rw{excavation} before
\emph{Howard Carter} and a treaty clip surfaces \rw{negotiation} before \emph{Egypt
and Israel}. In \emph{speech-content recovery} the readout recovers a spoken detail:
on a Kidney clip the specific organ, named only in the audio, becomes rank 1 in the
band, while the question-echoed \rw{transplant} does not count. In \emph{sound-source
identification} (Lion, Bird, Music, Train, Whip) the readout names the source
directly from the acoustics. The recurrence of non-English scripts at rank
1---\niaolei\ (birds), \emph{m\'usica} (music)---is the multilingual signature of
\cref{sec:f2}.

\begin{table}[t]
\centering
\small
\begin{tabular}{llccrr}
\toprule
Answer & Top readout & Layer & Depth & Rank & \# ws.\ hits \\
\midrule
Dallas & \rw{assassination} & L30 & 62\% & 1 & 56 \\
Kidney & \rw{kidney} & L17 & 35\% & 1 & 44 \\
Lion & \rw{lions} & L19 & 40\% & 1 & 37 \\
Bird & \rw{\niaolei} (birds) & L33 & 69\% & 1 & 12 \\
Music & \rw{m\'usica} & L18 & 38\% & 1 & 11 \\
Train & \rw{rail} & L19 & 40\% & 1 & 27 \\
Whip & \rw{\bian} (whip) & L32 & 67\% & 1 & 7 \\
\bottomrule
\end{tabular}
\caption{\textbf{Seven curated thinking-path cases.} For each we report the strongest
(lowest-rank) workspace-band readout of a task-relevant concept over the audio span,
the layer (of 48) where it is most stably rank 1, its depth, its rank among the
152{,}064-token vocabulary, and the number of workspace cells with the concept in the
top-8. For Dallas we list \rw{assassination} as the workspace-band scene concept, but
it is partly transcribed by the model's own ASR (\cref{sec:f1}). Kidney is
speech-content recall; the rest are sound-source identifications.}
\label{tab:cases}
\end{table}

\paragraph{It is the answer we are reading, not frequent words.} A lens can surface
generically frequent tokens, so we run the same readout looking for ten unrelated
placebo words (\emph{banana}, \emph{guitar}, \emph{planet}, \emph{umbrella},
\emph{tuesday}, \dots). Across the ten clips the true answer concept fills 1--34\% of
workspace cells, while the placebos fill $0.12\%$---essentially never. The single
visible exception is \emph{trumpet} on the whip clip ($1.3\%$), where a whip-crack
genuinely resembles a brass transient.

\paragraph{The signal is sparse across time.} Unlike text, where readable content
spreads over many tokens, only 4--30\% of audio positions carry the concept in the
workspace band (\eg 5 of 34 positions on the Kennedy clip, 10 of 37 on the
archaeology clip). A randomly chosen audio position often reads as noise. This is why
we aggregate over positions by taking the best rank rather than averaging, which
would wash the signal out.

\begin{figure}[t]
\centering
\includegraphics[width=0.72\textwidth]{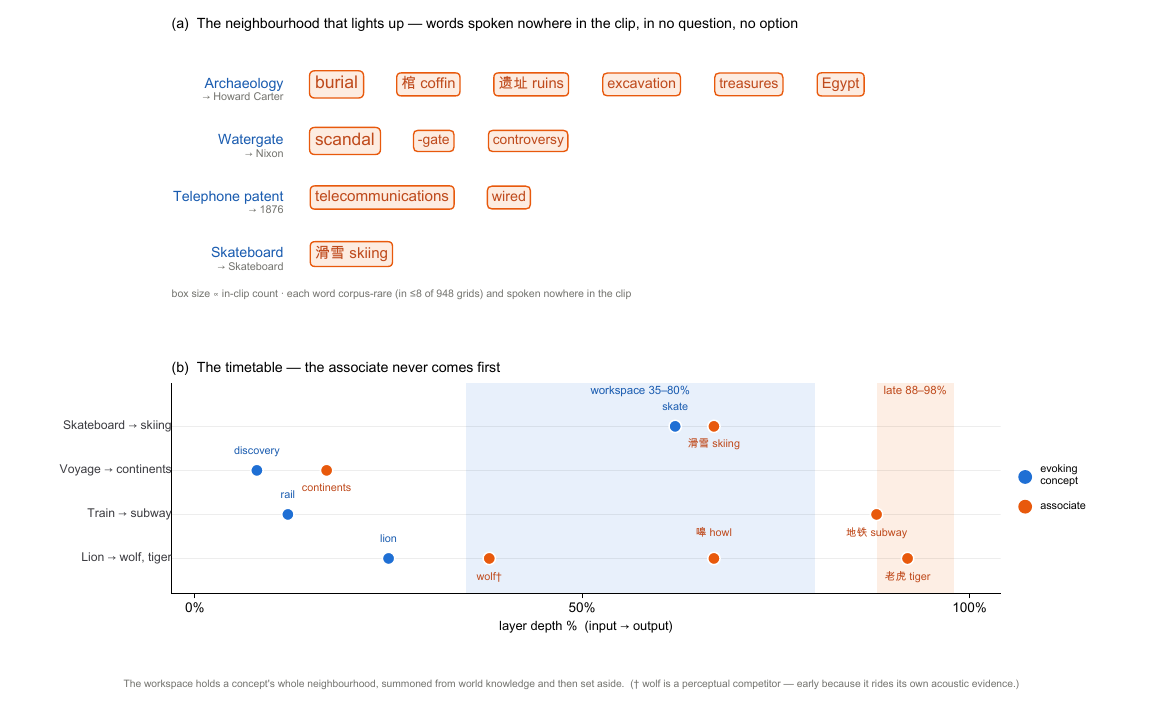}
\caption{\textbf{Watching the model free-associate.} \textbf{(top)} Spontaneous
neighbour words that light up at the audio positions---spoken nowhere in the clip, in
no question or option; corpus-rare. \textbf{(bottom)} Timetable: for each family, the
first readable layer of the evoking concept (blue) vs.\ each associate (orange); the
associate always follows its evoker.}
\label{fig:assoc}
\end{figure}

\paragraph{The workspace also holds the neighbourhood.} Beyond the task-relevant
concept, semantic neighbours of what the model hears light up spontaneously---spoken
nowhere in the clip, in neither question nor option---and are then set aside. On a
skateboard clip (a rolling sound; answer Skateboard), a single audio position read
down the layers shows the schedule: the raw sound is verbalized (\gundong,
\dq{rolling}), the category forms (\rw{skate}/\rw{skateboard}), and only then the
neighbour ignites (\huaxue, skiing, a sibling board-sport)---held for several layers
but never rank 1, then dropped (\cref{fig:assoc}). Across the families we could
verify, the associate always follows its evoker, and the farthest, evidence-free one
(\laohu, tiger) lands latest, $\sim$92\% depth. The richest cases are whole
neighbourhoods: the archaeology clip reconstructs a dig scene (\rw{burial},
\rw{coffin}, \rw{ruins}, \rw{excavation}, \rw{treasures}). A neighbour counts as
genuine only if absent from the clip's transcript on inspection and corpus-rare
($\leq$8 of 948 grids); literally spoken words are excluded. This is largely not an
unembedding-geometry artifact: because the lens projects onto $W_U$, a strengthening
evoker would lift its embedding neighbours for free, but five of six associates fall
outside their evoker's top-100 cosine neighbours, and the one near pair
(\emph{skiing}, the 7th neighbour of \emph{skate}) still ignites strictly after its
evoker, stays below rank 1, and is dropped---a schedule static similarity does not
encode. This is the audio analogue of spreading activation \citep{collins1975spreading}, read
layer by layer with depth playing the role of time.

\subsection{Finding 2: The Thinking Space Is Multilingual}
\label{sec:f2}

\begin{figure}[t]
\centering
\includegraphics[width=0.66\textwidth]{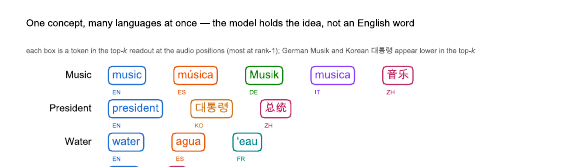}
\caption{\textbf{The workspace is multilingual.} The same concept surfaces across
languages at the audio positions (most at rank 1). The model holds a
language-agnostic concept rendered into whichever token sits nearest, not a memorized
English string.}
\label{fig:multiling}
\end{figure}

Is the model re-transcribing the sound, or holding a concept? Though clip and prompt
are entirely in English, one concept appears in several languages at once. On a music
clip the idea of music appears in the workspace band as English \rw{music}, Chinese
\yinyue, Spanish \rw{m\'usica}, German \rw{Musik} and Italian \rw{musica} at once
(\cref{fig:multiling}); \emph{president} surfaces as English, Chinese \zongtong\ and
Korean; \emph{water}, on the Nixon clip, as English, Spanish \rw{agua} and French
\rw{'eau}. Tallying the top-1 readout across all audio-region workspace cells
(863{,}770 cells over 948 clips, restricted to real-word readouts), the inner
vocabulary is $52.6\%$ English, $38.5\%$ Chinese and $\leq 4.3\%$ others---so more
than a third of leading readouts are Chinese, on English inputs.

This is not the lens's known bias toward frequent CJK tokens. A frequency control
shows the Chinese form is clip-specific: \yinyue\ is rank 1 on 30\% of music clips
but 6\% of the 610 non-music clips (a $5\times$ lift), and Chinese \emph{bird} forms
show $9$--$20\times$ lift on birdsong, where a globally frequent token would lift near
$1$.

\subsection{Finding 3: When Speech Carries Extra Information, the Audio Mind
Registers It}
\label{sec:f3}

\begin{figure}[t]
\centering
\includegraphics[width=0.72\textwidth]{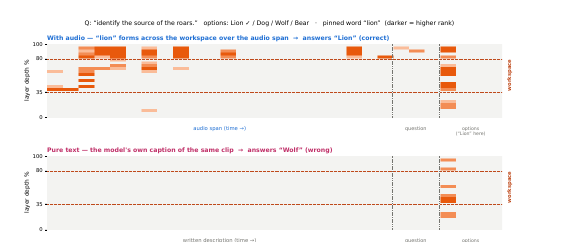}
\caption{\textbf{Two minds, one clip.} The roaring clip read two ways, with the
concept \emph{lion} pinned (darker $=$ better rank). \textbf{Top} (audio):
\rw{lion} lights up across the workspace over the audio span; the model answers
\emph{Lion}. \textbf{Bottom} (the model's own emotion-free caption): \rw{lion}
appears almost nowhere but the printed option, and the model answers \emph{Wolf}.}
\label{fig:twominds}
\end{figure}

The sound also carries \emph{how} something was said---an emotion, a speaker's role,
an acoustic source---which no transcript preserves. Reading the same clip twice, as
audio and as the model's own emotion-free caption of it, shows where that content
lives. On clips where the two minds disagree, the audio mind forms the true,
speech-specific concept and answers correctly while the caption mind does not: a roar
reads \rw{lion} (the caption run answers \emph{wolf}; \cref{fig:twominds}); a clip
reads \rw{priest} where the caption hears only \dq{father}; a sarcastic voice reads
the inferred affect \rw{frustration}, named nowhere in the prompt and nowhere in the
caption. Quantifying the disagreement clips, the audio mind forms the correct speaker
role on $88.9\%$ of them versus $70.4\%$ for the caption mind ($n{=}27$), and $33$ of
the $39$ affects it reads are absent from the caption mind's readout. The behavioral
shadow of this needs no lens---over the 1000-clip MMAU-mini, answering from the audio
beats answering from the caption on every track, by $7.6$ points overall and $11.1$ on
the sound track---but the point is not the score: it is that the concept the answer
depends on is formed in one mind and not in the other. This also distinguishes the
effect from prior attribute-tracking \citep{yang2025audiolens}: the affect is a finer,
inferred state named nowhere in the transcript, contrasted against a caption mind that
does not form it.

\paragraph{The opposite regime, for calibration.} Hearing does not always help, and we
state the boundary rather than hide it. On spoken TriviaQA the answer is a stored fact
and the spoken clip carries words identical to the written question, so the sound adds
nothing the text does not already supply, and the model does at least as well reading
as listening. Reading both minds on the items it gets right from text and wrong from
speech ($n{=}87$), the answer concept appears in the text mind's workspace on $83\%$
of them versus $62\%$ for the speech mind---a partial retrieval loss, mixed with items
where the concept does reach rank 1 in the speech workspace and the spoken output
still garbles it. We therefore treat this task as the boundary of the effect, not as
a clean mechanism.

\subsection{Finding 4: The Thought Lives in the Middle, and Forms in Depth Order}
\label{sec:f4}

The findings above read the workspace band by assumption. We now show that the
audio-driven signal is most cleanly decodable in the middle band, turns on early, and
is causally used---committed before the output---not merely present.

\paragraph{Is the readout driven by the sound?} Under the audio-swap control
(question and options byte-identical; only the waveform real / mismatched / silence),
the correct answer is read far more strongly under real audio ($40.0\%$ balanced
accuracy) than under silence ($21.8\%$, at chance, where a majority guess scores
$25\%$; paired
McNemar $p{=}4.7\times10^{-5}$). Because silence has fewer audio positions and our
best-rank metric takes a minimum over positions, we also rest the claim on the
position-matched real-vs-mismatch contrast, where the two conditions have nearly
equal audio-position counts ($47.9$ vs.\ $43.6$ mean positions): real ($40.0\%$)
still beats mismatch ($32.2\%$) in the workspace band ($p{=}0.015$). Threshold-free,
the median rank of the correct answer among the $152{,}064$ vocabulary tokens degrades
with signal quality in the same order: $\#2{,}297$, $\#5{,}040$, $\#26{,}241$.

\paragraph{Where: the audio signal separates from the text prior in the middle band.}
Sweeping the readout band by band (\cref{tab:bands}), the
real$-$silence gap---the audio content the text prior alone cannot supply---is
negligible in the early sensory band (real $\approx$ silence, both explained by the
text prior; gap $+3.0$, $p{=}0.78$), largest in the workspace band (silence collapses
to chance there, $38.0 \rightarrow 21.8$, while real holds; gap $+18.2$,
$p{=}4.7\times10^{-5}$), and intermediate in the motor band ($+14.2$,
$p{=}1.5\times10^{-3}$). We are precise about what this shows and report both
contrasts: the workspace is where the audio signal is most cleanly \emph{separable}
from the text prior---the silence baseline falls to chance there and nowhere
else---not where its raw magnitude is greatest, since under the position-matched
real$-$mismatch contrast magnitude keeps growing toward the output.

\begin{table}[t]
\centering
\small
\begin{tabular}{lccccc}
\toprule
Band & Real & Sil. & Mism. & R$-$S & R$-$M \\
\midrule
Sensory   & 41.0 & 38.0 & 31.8 & $+3.0$  & $+9.2$ \\
Workspace & 40.0 & 21.8 & 32.2 & $+18.2$ & $+7.8$ \\
Motor     & 48.9 & 34.7 & 26.3 & $+14.2$ & $+22.6$ \\
\bottomrule
\end{tabular}
\caption{\textbf{Balanced four-way readout accuracy (\%) by functional band.} R$-$S
(real$-$silence) tests separation from the text prior---null in the sensory band
($p{=}0.78$), largest and highly significant in the workspace band
($p{=}4.7\times10^{-5}$) where silence alone collapses to chance, intermediate in
motor ($p{=}1.5\times10^{-3}$). R$-$M (real$-$mismatch, position-matched) is
significant in the workspace ($p{=}0.015$) but grows toward motor
($p{<}10^{-3}$), so raw magnitude keeps sharpening toward the output while
separability from the prior peaks mid-stack. $p$: paired two-sided McNemar.}
\label{tab:bands}
\end{table}

\begin{figure}[t]
\centering
\includegraphics[width=0.66\textwidth]{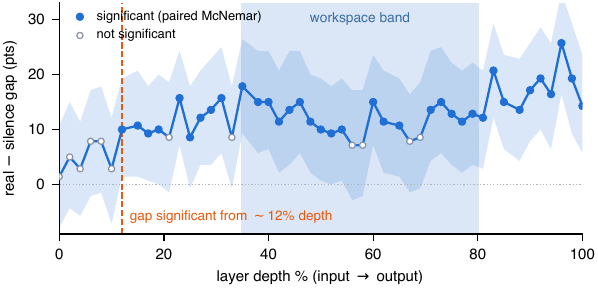}
\caption{\textbf{The audio-driven gap over the text prior, layer by layer.} Balanced
four-way readout accuracy with real audio minus the same quantity with silence, with
a 95\% bootstrap band; filled markers are depths where the paired test is
significant. The gap is null at the input, becomes significant at $\sim$12\% depth,
and stays open, densest inside the workspace band.}
\label{fig:onset}
\end{figure}

\paragraph{When: the signal turns on early and does not fade.} Reading the
real$-$silence gap layer by layer (\cref{fig:onset}), the two conditions are
statistically indistinguishable through the first several layers---the encoder has
deposited acoustic features, but nothing the sound adds is yet answer-relevant---and
then, a little more than a tenth of the way up, the gap opens and stays open: 37 of
the 49 readout depths (the embedding output plus the 48 layers) reach $p<0.05$
(paired McNemar, one-sided, as the audio-driven direction is predicted a priori), 35
surviving Benjamini--Hochberg control, concentrated in the workspace band (18 of 22,
versus 9 of the 17 sensory layers). This transition is not the model beginning to
\emph{process} the audio---the encoder writes acoustic features into the residual
stream from the first layer---but the depth at which the sound's content becomes
verbalizable and decision-relevant.

\paragraph{Depth order on a single clip.} The staging is visible on individual clips,
not only on average. On the Kennedy clip (\cref{fig:chain}) the audio-inferred
concepts appear in order: the event (\rw{death}, \rw{assassination}) is readable from
$\sim$8\% depth and holds through the workspace, while the victim's name
(\rw{Kennedy}) surfaces only in the last layers and the answer commits at the output.
The Nixon clip repeats the pattern (\cref{fig:nixon}): \rw{scandal} and the
reconstructed \rw{water}${+}$\rw{gate} early, the role \rw{president} in the
workspace, the answer by the middle layers.

\paragraph{Causally used, and committed before the late layers.} Decodability need not
mean use. On clips the model gets right with audio and wrong once the audio is zeroed,
patching the real-audio activations back at one band restores the correct answer:
$10/10$ clips flip back with a sensory-band patch (95.7\% of the clean$-$corrupt logit
gap recovered) and $9/10$ with a workspace-band patch (89.1\%), while patching only
the motor band restores nothing ($0/10$, 4.8\%). So the audio content is causally used
and committed before the motor band. Because an early patch propagates downstream,
this localizes coarsely: it shows the content is used and committed before the last
fifth of the network, not that the workspace band alone is responsible---a
workspace-band patch at the \emph{text} positions, which have already attended to the
audio, also restores the answer ($9/10$).

\paragraph{Not one checkpoint's quirk.} Repeating the band-localized audio-swap on an
architecturally different model---Qwen2.5-Omni-7B \citep{xu2025qwen25omni}, a dense
28-layer Thinker versus our sparse mixture-of-experts---reproduces the signature: the
real$-$silence gap is small in the sensory band and large and highly significant in
the workspace band ($p<10^{-3}$), the same sensory-null-to-workspace-live ordering as
\cref{tab:bands}.

\subsection{Finding 5: The Answer Is Distributed, and Hallucinations Are Born Early}
\label{sec:f5}

\begin{figure}[t]
\centering
\includegraphics[width=0.86\textwidth]{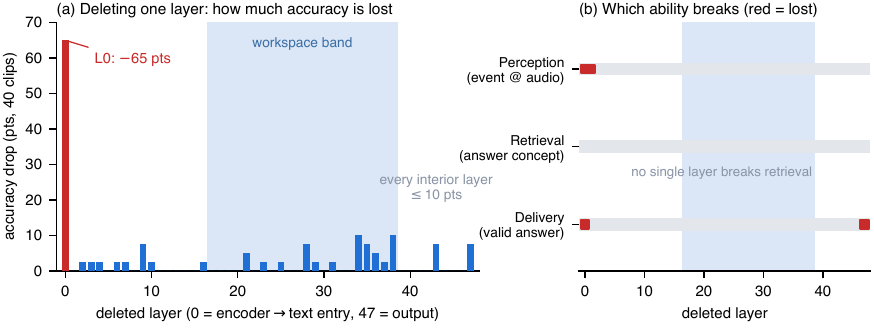}
\caption{\textbf{Deleting one layer at a time.} \textbf{(a)} Accuracy drop from
bypassing each layer, over 40 clips. Only the entry layer L0 is critical
($-65$ points); every interior layer costs at most 10. \textbf{(b)} Which ability is
lost. Perception fails only for L0--L1, delivery only for L47, and retrieval for no
single layer---listening and delivery are localized to the two ends, retrieval is
distributed across the interior.}
\label{fig:ablation}
\end{figure}

Reading is correlational; deleting is causal. Bypassing one layer at a time (an
identity skip) and measuring accuracy over 40 clips gives a clean functional map
(\cref{fig:ablation}). Only L0---the encoder-to-text entry, where the audio is read
in---is clearly critical: deleting it costs 65 points of accuracy. Every interior
layer costs at most 10 points, and the band means excluding L0 are all within one
point of zero (sensory $-0.2$, workspace $+0.7$, motor $+1.0$). The answer is therefore not carried by
any one interior layer; the network routes around whichever one is removed.

Breaking accuracy is one thing, and which \emph{ability} breaks is another. On the
Kennedy clip we track three separable functions per deleted layer: perception (does
the sound-inferred event \rw{assassination} still form at the audio positions?),
retrieval (does the recalled answer concept still surface?), and delivery (does the
model still emit a valid answer?). Only deleting L0--L1 stops the event forming from
the sound; only deleting L47 breaks the output; and no single layer, deleted, stops
the answer concept from surfacing. The pipeline is thus: read the sound in (L0--L1),
hold and recall the answer across the interior, deliver it out (L47)---and only the
two ends are irreplaceable.

\begin{figure}[t]
\centering
\includegraphics[width=0.42\textwidth]{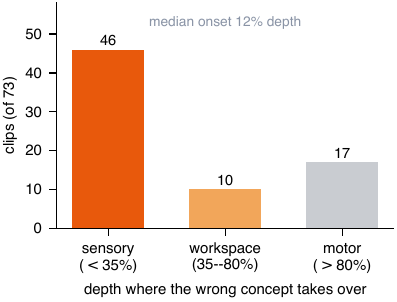}
\caption{\textbf{Where a text-mode hallucination is born.} Depth at which the wrong
concept first takes over the caption-fed run, on 73 clips the model answers correctly
from the audio. Most wrong answers are early commitments, formed inside the sensory
band and carried forward unchanged.}
\label{fig:halluc}
\end{figure}

\paragraph{Where a hallucination is born.} The same readout localizes errors. Take the
clips the model answers correctly from the real audio but wrongly from its own
caption: the caption dropped the decisive acoustic detail, so the text-only mind falls
back on a plausible neighbour. Across 73 such clips we trace the depth at which the
wrong concept first takes over the text mind (\cref{fig:halluc}). It is usually an
early commitment---median onset 12\% depth, 58.9\% formed by 20\% depth, 46 of 73
inside the sensory band---and 84.9\% of the wrong answers form before the output
layer rather than slipping at the end. A text-mode hallucination is thus typically not
a late slip: the model locks onto a plausible-but-wrong concept at roughly the same
depth where correct answers first emerge (\cref{sec:f4}) and never revises it. Reading
the two minds side by side both detects the hallucination---the audio mind holds the
right concept where the text mind holds the wrong one---and shows where it entered.

\section{Conclusion}

We asked whether the verbalizable middle-layer global workspace found in text, code,
and image models also exists in an audio LLM. Reading the Thinker of a base
Qwen3-Omni model with a logit lens at the audio positions, we find evidence that it
does: the answer-relevant concept becomes readable before any token is emitted, it is
driven by the sound (waveform-swap control) and causally used before the output
(activation patching), and the readable content is conceptual rather than a
re-transcription---multilingual, paralinguistic, and free-associating. Deleting layers
one at a time adds the functional map: listening and delivery are localized to the two
ends of the stack, while retrieval is distributed across the interior. This is
precisely the regime a chain-of-thought monitor cannot see, since stated reasoning need
not be produced and can be unfaithful.

The account is deliberately scoped. The logit lens is a cheap proxy---relative to a
faithful lens such as the Jacobian lens of \citet{gurnee2026workspace} it should, if
anything, under-detect, and early-layer readouts are noisy by construction, which is
why our claims concern the middle band; the patching test likewise localizes coarsely,
showing the audio content is committed before the motor band rather than pinning it to
the workspace alone. We use \dq{workspace} and \dq{access} in the functional,
information-posted-for-report sense of \citet{baars1988cognitive} and \citet{dehaene2001workspace}, and
make no claim about subjective experience. The natural next steps are a corpus-scale
replication under a faithful lens, larger forced-alignment sets, and a test of whether
safety-relevant decisions---tool calls, refusals, fabrication---are readable this way
before a speech agent acts.

Two uses would follow from that test. As a \emph{training} signal: reinforcement
learning is now the standard way to shape generative models
\citep{fan2025online,fan2025adaptive,gdi,lbc,yang2026batched}, and a workspace readout
offers a process-level target where a scalar reward sees only the final answer. As a
\emph{monitor}: deployed multimodal systems are latency- and memory-bound, which is why
so much work goes into adaptive inference and pruning
\citep{prance,li2026spvla,elegantvla}; a logit-lens probe is by comparison almost free,
one matrix product per position read.

\bibliographystyle{plainnat}
\bibliography{main}

\end{document}